\documentclass[quantumrep,article,accept,pdftex,moreauthors]{Definitions/mdpi} 
\firstpage{1} 
\pubvolume{8}
\issuenum{3}
\articlenumber{74}
\pubyear{2026}
\copyrightyear{2026}
\externaleditor{Chao Zheng and Jim Freericks} 
\datereceived{30 June 2026} 
\daterevised{27 July 2026} 
\dateaccepted{29 July 2026} 
\datepublished{31 July 2026} 
\pdfoutput=1 

\usepackage[english]{babel}
\usepackage[normalem]{ulem}

\usepackage{graphicx}
\usepackage{amsmath,amssymb,mathrsfs,esint}
\usepackage{mathtools}
\usepackage{physics2}
\usephysicsmodule{ab, braket}
\let\Re\relax
\let\Im\relax
\DeclareMathOperator{\Tr}{Tr}
\DeclareMathOperator{\sinc}{sinc}
\DeclareMathOperator{\Re}{Re}
\DeclareMathOperator{\Im}{Im}

\usepackage{subcaption}

\usepackage{bm}
\usepackage{relsize}

\usepackage[normalem]{ulem}

\usepackage{tikz}

\DeclareUnicodeCharacter{00A0}{~}

\Title{Influence of fast and slow laser phase noise on the fidelity of the M\o lmer-S\o rensen trapped-ion gate}

\Author{Nikita Semenin $^{1}$*\orcidA{}, Ksenia Khabarova $^{1,2}$ and Nikolay Kolachevsky $^{1,2}$}

\AuthorNames{Nikita Semenin, Ksenia Khabarova and Nikolay Kolachevsky}

\address{%
$^{1}$ \quad P.N. Lebedev Physical Institute of the Russian Academy of Sciences, Moscow 119991, Russia; habarovaky@lebedev.ru (K.K.); kolachevsky@lebedev.ru (N.K.)\\
$^{2}$ \quad Russian Quantum Center, Skolkovo, Moscow 121205, Russia}

\corres{Correspondence: semeninnv@gmail.com}

\abstract{High-fidelity two-qubit entangling gates are essential for the realization of useful quantum algorithms on quantum processors. The M\o lmer-S\o rensen (MS) gate has become a common \textcolor{black}{choice} for trapped-ion quantum computing due to its resilience to ion temperature and its demonstrated record fidelities. However, the spectral impurity of the driving laser field impacts gate performance, with phase noise influencing the qubit dynamics through mechanisms operating on different timescales. In this work, we present a comprehensive theoretical analysis of laser phase noise in the MS gate, identifying two spectral ranges which influence the gate fidelity the most: "fast" noise at frequencies near the motional mode spectrum, and "slow" noise at frequencies on the order of the inverse gate time. We derive the noise Hamiltonians for two common laser beam geometries and obtain analytical expressions for the average gate error in terms of the laser noise power spectral density and gate parameters. For slowly varying noise spectra, we provide simplified error estimates.
In addition, we validate our findings against previously published numerical simulations.
}

\keyword{Trapped-ion quantum computing; Laser phase noise; MS gate}

\begin{document}




\section{Introduction}
The anticipated progress in development of large- and intermediate-scale quantum computers~\cite{abughanem2025ibm, google2025quantum, ransford2025helios} relies on the ability to perform high-fidelity gates in many-qubit quantum registers. Although there are algorithmic methods of compensating for imperfect gates, such as error correction~\cite{nielsen_quantum_2010} and error mitigation~\cite{cai2023quantum}, the computational and/or experimental overhead imposed by these methods may significantly reduce their advantage and limit their applicability. Thus, it remains an important task to identify, study and suppress the various physical mechanisms that lead to gate errors.

Among all physical platforms for quantum computation, trapped ions continue to prove their efficacy, having already demonstrated record one- and two-qubit gate fidelities~\cite{loschnauer_scalable_2025} and quantum volume~\cite{quantinuum_QV}, with recent larger-scale high-connectivity systems approaching 100 qubits~\cite{ransford2025helios}. The long-range strong Coulomb interaction between the ions, as well as the invention of the quantum charge-coupled device (QCCD) architecture~\cite{kielpinski2002qccd}, allow efficient and flexible scaling and high-fidelity entanglement~\cite{pino2021demonstration, moses_race-track_2023}.

\begin{figure}[h]
    \centering
    \includegraphics[width=0.7\linewidth]{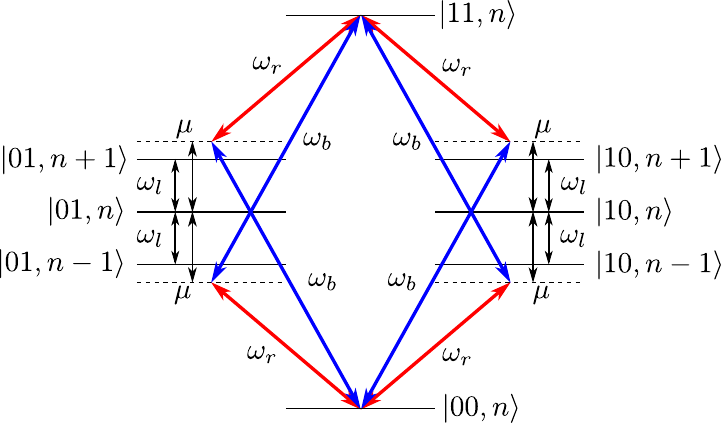}
    \caption{The MS gate level scheme. The bichromatic field ($\omega_b,\omega_r$) creates 4 Raman paths between states~$\ket{00, n}\leftrightarrow\ket{11, n}$. The intermediate states are the blue and red motional sidebands of the~$l$-th mode with frequency~$\omega_l$}
    \label{fig:MS_level}
\end{figure}

One of the major challenges, characteristic of any platform, is the fidelity of two-qubit entangling gates. Any universal quantum computer requires at least one such gate to be able to perform any unitary transformation of the whole register with a given accuracy in finite time~\cite{nielsen_quantum_2010}. With trapped ions in particular, one common choice is the M\o lmer-S\o rensen (MS) gate~\cite{sorensen_quantum_1999, sorensen_entanglement_2000}. In the original proposal, it invokes a \emph{bichromatic} laser field, which contains two frequency components, symmetrically detuned (by~$\pm\mu$) from the qubit carrier transition, called the \emph{red} and \emph{blue} components (Fig.~\ref{fig:MS_level}). The detuning is chosen close to one of the motional modes of the ionic chain in the trap, creating 4 Raman-like stimulated transitions, with the intermediate states being the motional sidebands. Due to interference between the transitions, the effective frequency of the Rabi cycle between states~$\ket{00}$ and~$\ket{11}$ will be independent on the motional number~$n$. This is the main reason the MS gate is used extensively in many setups that show high two-qubit fidelities~\cite{schafer_fast_2018, loschnauer_scalable_2025, hughes2025trapped}. Although the original idea assumed that the transition~$\ket{0}\rightarrow\ket{1}$ is an optical transition capable of being excited directly by each laser beam, the description is valid for hyperfine qubits driven by two Raman laser beams~\cite{debnath_demonstration_2016, chen_benchmarking_2024}, or even for laser-free setups~\cite{loschnauer_scalable_2025,ospelkaus_microwave_2011,srinivas_high-fidelity_2021}.

\begin{figure}[h]
    \centering
    \includegraphics[width=0.4\linewidth]{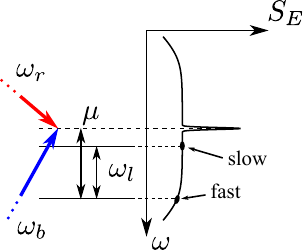}
    \caption{The \textcolor{black}{influence} of laser \textcolor{black}{phase} noise on the MS gate. Parasitic interactions arise from the overlap of the laser field power spectral density (PSD)~$S_E(\omega)$ with the closest states of the system, \textcolor{black}{outlining} the "slow" and "fast" parts \textcolor{black}{(see main text)}}
    \label{fig:MS_level_noise}
\end{figure}

\textcolor{black}{
By construction of the MS gate, it is clear that its fidelity will be sensitive to phase fluctuations of the fields driving it, since the Raman process described above requires coherence between the two bichromatic components. All those components for all ions are usually generated by a single laser source, which then has its radiation split into several beams and then modulated with various optoelectronics. Therefore, the intrinsic phase noise of this source will transfer onto every field present in the gate. This will lead to the bichromatic components depicted in Fig.~\ref{fig:MS_level} having acquired an identical spectrum~$S_E(\omega)$ in terms of the electric field power spectral density (PSD), which is centered at the intermediate point between the red and blue parts (see Fig.~\ref{fig:MS_level_noise}) and is related to the phase noise PSD~\cite{riehle_frequency_2006}. This spectrum will, in general, overlap with the levels of the quantum system closest to the central peak, causing non-resonant parasitic transitions to and from these levels. This is similar to how a small perturbation with a continuous spectrum causes transitions between discrete levels with a rate obtained by applying the Fermi's golden rule.}

\textcolor{black}{There are two most significant interactions that influence the MS gate error. The first one is the interaction with the carrier transition~$\ket{0, n}\rightarrow\ket{1, n}$. It is caused by the Fourier components of the phase noise that have frequencies close to the detuning~$\mu$. Since this detuning is, in turn, close to the motional frequencies of the ions (hundreds of kHz to several MHz), the timescales at which this noise is present are on the order of the oscillation period in the trap ($\sim 1\mu$s and less). At these high frequencies the main physical source of phase noise is the servo loop which is used to lock the laser to a stable reference, such as a high-finesse Fabry-P\'erot cavity~\cite{kolachevsky2011low, Alnis2008Subhertz, schmid2019simple, Zalivako_2020}. At frequencies around the servo bandwidth (which also lies in the MHz range~\cite{nakav_effect_2023, Senko2022limits}), the noise of the free-running laser is amplified instead of being suppressed, causing the so-called servo bumps~\cite{nakav_effect_2023}. These bumps have, therefore, a sizable effect on the fidelity of the MS gate, if they have a sufficient spectral overlap with the carrier transition. Another source of high-frequency errors are the relaxation oscillations of the gain medium. For solid state lasers, the relaxation oscillation peak in the phase noise spectrum lies in the sub-MHz range~\cite{Senko2022limits, lu2011experimental, zhang2012compact, galzerano2007single} and is thus also able to contribute to the gate error. Due to the small timescale of the described interaction, we call this noise the "fast" noise, since usually the gate time is much larger than the trap oscillation period.}

\textcolor{black}{The second interaction is the interaction with the first motional sidebands~$\ket{0, n}\rightarrow\ket{1, n\pm 1}$ The relevant noise components in this case are on the order of~$\mu-\omega_l$ and lie in the kHz to tens of kHz range. This frequency is approximately the inverse gate time, so the timescale for this type of noise is $0.1\div 1$ ms. Moreover, frequency deviations of the laser that are slower than this timescale but have a large magnitude ($\lesssim\mu-\omega_l$) can lead to significant errors due to the bichromatic components being close to the resonance with the sideband transition. This "slow" noise (compared to the "fast" noise described above) can be caused by acoustic or thermal effects, and in the case of a cavity-stabilized laser, they are mainly tied to the mechanical properties of the cavity itself, owing to the large servo gain at such frequencies. These sources yield a~$1/f$ frequency noise PSD~\cite{numata2004thermal, kessler2011thermal}. Another source with the same~$1/f$ PSD is the flicker noise, which is present in almost any electronic device, including the servo loop which locks the laser to the cavity. Technical and flicker noise together provide the low-frequency part of laser noise with a cutoff frequency of~$10\div100\ \text{Hz}$~\cite{milani2017multiple, westergaard2010strontium}, above which it is overtaken by white noise with a flat PSD~\cite{milani2017multiple, westergaard2010strontium, Senko2022limits}.}

To our knowledge, there seems to be limited attention to the \textcolor{black}{two described types of} noise in the literature. Despite the existence of various methods of reducing the influence of phase noise, such as using decoherence-free subspaces~\cite{Lidar1998decoherence, barenco1997stabilization, Zanardi1997noiseless, zanardi2000stabilizing, Duan1997preserving}, as well as dynamical decoupling techniques~\cite{ban1998photon,Viola1999dynamical, manovitz_fast_2017,zalivako_continuous_2023}, they only deal with pure dephasing, whereas the noise described above causes parasitic \textcolor{black}{interactions}, which these methods fail to adequately address. Fast laser noise of this type was studied in~\cite{nakav_effect_2023}, however, the authors estimated the error induced by this noise from numerical simulations, without calculating the error analytically. Time-varying noise of the laser phase was analyzed in~\cite{kang_designing_2023}, but the analysis there was confined to a specific laser beam geometry, and only fluctuations in the beam optical paths were considered, not the \textcolor{black}{inherent phase noise of the common laser source which produces all individual addressing fields}.

Here, we perform a theoretical analysis of \textcolor{black}{intrinsic phase noise of the laser} in the MS gate. We derive the noise terms in the Hamiltonian for both spectral ranges (fast and slow) and for all common beam geometries. 
\textcolor{black}{We show that the average gate error is equal to the convolution between the phase or frequency noise PSD (depending on whether fast or slow noise is considered, respectively) and the so-called \emph{filter function} of the gate, with the latter only depending on the gate parameters.}
We obtain analytical expressions for the filter functions,
as well as give simplified estimates for the error in trivial cases. Note that, despite our theory assuming laser radiation, as in the original description of the MS gate, it is applicable to all alternative gate setups mentioned earlier.
\textcolor{black}{Our findings complement previously published literature in two main ways. The first original result are the analytical expressions for the filter functions in both common beam geometries and both fast and slow spectral ranges. These filter functions expand on the filter function theory for trapped ions developed in~\cite{kang_designing_2023}, where a different noise source was studied. The second original result are the simplified estimations of the average error, which not only confirm the numerical simulations from~\cite{nakav_effect_2023}, but also can be used as a rule of thumb for calculating the requirements on the spectral purity of the laser given the needed gate error.}

In Section~\ref{sec:MS_gate_hamiltonians}, we derive the MS gate Hamiltonians for two commonly used laser beam configurations (see below) and define the noise terms contributing to fast and slow noise.
Sections~\ref{sec:fast_noise} and~\ref{sec:slow_noise} are devoted to calculating the filter functions for the fast and slow parts of the noise, respectively. Section~\ref{sec:Results} presents the visual graphs of the calculated filter functions, as well as estimates of the total gate error in the simplest cases. We discuss our results and compare them to the findings of previously published works in Section~\ref{sec:discussion}.

\section{Materials and Methods}
\subsection{Phase-sensitive and phase-insensitive MS gate Hamiltonians}
\label{sec:MS_gate_hamiltonians}
\begin{figure}[h]
\vspace{5mm}
    \centering
    \includegraphics[width=0.7\linewidth]{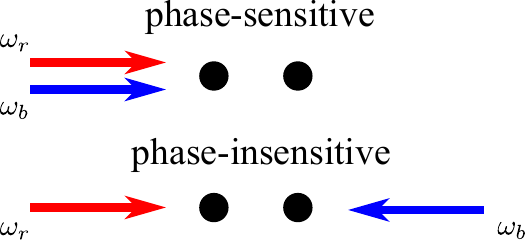}
    \caption{The phase-sensitive and phase-insensitive laser beam geometries. In this example the axial motional modes of the ionic chain are excited}
    \label{fig:MS_setups_sensitive_insensitive}
\end{figure}

We start by deriving the MS gate Hamiltonians for the two most widely used laser field configurations: \emph{phase-sensitive} and \emph{phase-insensitive}~\cite{lee_phase_2005}. In the phase-sensitive geometry, the bichromatic beam components are co-propagating, whereas in the phase-insensitive geometry they are counter-propagating (Fig.~\ref{fig:MS_setups_sensitive_insensitive}).
The names of the configurations come from the different sensitivity of the entangling term in the evolution operator to fluctuations in the beams' optical paths~\cite{lee_phase_2005}. In both cases, we assume the intensities of the bichromatic components to be equal at the ion's position. The Hamiltonian for both configurations will have the following form:
\begin{multline}
\widehat{H}_{\mathrm{MS}}(t) = \sum\limits_j\dfrac{\Omega_j}{2}\left\lbrace\left[\sigma_+^{(j)}\exp\left(\pm i\sum\limits_l \eta_{lj}(\hat{a}_le^{-i\omega_l t} + \hat{a}^{\dagger}_le^{i\omega_l t})\right)e^{i(\phi_b^{(j)}-\mu t)} + \mathrm{h.c.}\right]\right. \\\allowdisplaybreaks
\left. + 
\left[\sigma_+^{(j)}\exp\left(\pm i\sum\limits_l \eta_{lj}(\hat{a}_le^{-i\omega_l t} + \hat{a}^{\dagger}_le^{i\omega_l t})\right)e^{i(\phi_r^{(j)}+\mu t)} + \mathrm{h.c.}\right]
\right\rbrace,
\end{multline}
where $\Omega_j$ is the Rabi frequency of the bichromatic field (the index~$j$ denotes the~$j^{\mathrm{th}}$ ion), $\sigma_+^{(j)}$ is the raising Pauli matrix, $\eta_{lj}$ the Lamb-Dicke parameter of the~$l^{\mathrm{th}}$ motional mode with frequency~$\omega_l$ and ladder operators~$\hat{a}_l, \hat{a}^{\dagger}_l$. The phases of the red and blue field components are~$\phi_r^{(j)}$ and~$\phi_b^{(j)}$ respectively, and their detuning relative to the carrier transition is symmetric and equal to~$\mu$. Note that this expression is valid for both purely optical transitions and hyperfine Raman transitions, as long as the Lamb-Dicke parameters are defined properly~\cite{leibfried_quantum_2003}. The signs~$\pm$ are picked according to the specific laser field configuration, see below.
\subsubsection{Phase-sensitive configuration} In this configuration, both signs are chosen "$+$". In the Lamb-Dicke regime, the full Hamiltonian can be reduced to
\begin{equation}
\label{eq:MS_hamiltonian_sensitive}
\widehat{H}_{\mathrm{MS}}^{(\mathrm{S})}(t)\approx \sum\limits_j \Omega_j \sigma_{\phi_\perp}^{(j)}\cos(\mu t+\phi_M^{(j)})
-\sum\limits_{l,j} \eta_{lj}\,\Omega_j \sigma_{\phi}^{(j)}\cos(\mu t+\phi_M^{(j)})(\hat{a}_le^{-i\omega_l t} + \hat{a}^{\dagger}_le^{i\omega_l t}),
\end{equation}
where the "rotated" Pauli matrices are
\begin{equation}
\begin{aligned}
\sigma_{\phi_\perp}^{(j)}&=\sigma_{x}^{(j)}\cos\phi_S^{(j)}-\sigma_{y}^{(j)}\sin\phi_S^{(j)},\\\allowdisplaybreaks
\sigma_{\phi}^{(j)}&=\sigma_{x}^{(j)}\sin\phi_S^{(j)}+\sigma_{y}^{(j)}\cos\phi_S^{(j)},
\end{aligned}
\end{equation}
and the motional and spin phases~$\phi_M^{(j)},\phi_S^{(j)}$ are defined as
\begin{equation}
\label{eq:phi_s_m}
\phi_{S,M}^{(j)} = (\phi_r^{(j)}\pm\phi_b^{(j)})/2.
\end{equation}
\subsubsection{Phase-insensitive configuration} In this configuration, one sign is~"$+$", while the other is~"$-$". Keeping the same definitions from above, the Hamiltonian in the Lamb-Dicke regime is
\begin{equation}
\label{eq:MS_hamiltonian_insensitive}
\widehat{H}_{\mathrm{MS}}^{(\mathrm{I})}(t)\approx \sum\limits_j \Omega_j \sigma_{\phi_\perp}^{(j)}\cos(\mu t+\phi_M^{(j)})
+\sum\limits_{l,j} \eta_{lj}\,\Omega_j \sigma_{\phi_\perp}^{(j)}\sin(\mu t+\phi_M^{(j)})(\hat{a}_le^{-i\omega_l t} + \hat{a}^{\dagger}_le^{i\omega_l t}).
\end{equation}
\subsubsection{Laser noise terms}
The noise terms in the above Hamiltonians come from fluctuations of the phases~$\phi_r^{(j)}$ and~$\phi_b^{(j)}$. If both bichromatic components come from the same source, which is usually the case~\cite{debnath_demonstration_2016, loschnauer_scalable_2025, Aksenov2023realizing,zalivako2025towards}, their fluctuations are identical to the fluctuations in the source. This means that, according to~\eqref{eq:phi_s_m}, only the phase $\phi_S^{(j)}$ is affected:
\begin{equation}
\delta\phi_S^{(j)}(t) = \delta\phi(t),\quad\delta\phi_M^{(j)}(t) = 0,
\end{equation}
where~$\delta\phi(t)$ is the phase noise of the laser source. These fluctuations, in turn, lead to the following fluctuations in the spin operators:
\begin{equation}
\begin{aligned}
\sigma_{\phi}^{(j)}(t)&\approx \sigma_{\phi}^{(j)}+\sigma_{\phi_\perp}^{(j)}\delta\phi(t),\\\allowdisplaybreaks
\sigma_{\phi_\perp}^{(j)}(t)&\approx \sigma_{\phi_\perp}^{(j)}-\sigma_{\phi}^{(j)}\delta\phi(t).
\end{aligned}
\end{equation}
Inserting these expressions into~\eqref{eq:MS_hamiltonian_sensitive} and~\eqref{eq:MS_hamiltonian_insensitive}, we obtain the noise terms for the phase-sensitive configuration:
\begin{multline}
\label{eq:noise_hamiltonian_sensitive}
\widehat{V}^{(\mathrm{S})}(t)=-\delta\phi(t)\left[\sum\limits_j \Omega_j \sigma_{\phi}^{(j)}\cos(\mu t+\phi_M^{(j)})\right.\\\allowdisplaybreaks
\left.+\sum\limits_{l,j} \eta_{lj}\,\Omega_j \sigma_{\phi_\perp}^{(j)}\cos(\mu t+\phi_M^{(j)})(\hat{a}_le^{-i\omega_l t} + \hat{a}^{\dagger}_le^{i\omega_l t})\right],
\end{multline}
and for the phase-insensitive configuration:
\begin{multline}
\label{eq:noise_hamiltonian_insensitive}
\widehat{V}^{(\mathrm{I})}(t)=-\delta\phi(t)\left[\sum\limits_j \Omega_j \sigma_{\phi}^{(j)}\cos(\mu t+\phi_M^{(j)})\right.\\\allowdisplaybreaks
\left.+\sum\limits_{l,j} \eta_{lj}\,\Omega_j \sigma_{\phi}^{(j)}\sin(\mu t+\phi_M^{(j)})(\hat{a}_le^{-i\omega_l t} + \hat{a}^{\dagger}_le^{i\omega_l t})\right].
\end{multline}
The two sums in square brackets for each configuration represent the fast and slow parts of the noise, respectively, due to the reasons provided in the introduction. Furthermore, the separation between the relevant spectral ranges of the noise terms allows their independent treatment, since their spectral overlap is insignificant.
\subsubsection{Mean noisy gate fidelity}
In a general qubit system, the Hamiltonian of a noisy gate can be written as
\begin{equation}
\widehat{H}(t)=\widehat{H}_0(t)+\widehat{V}(t),
\end{equation}
where~$\widehat{H}_0(t)$ is the ideal Hamiltonian and~$\widehat{V}(t)$ is the noise. The full evolution operator can be represented in the following way:
\begin{equation}
\label{eq:evolution_operator_perturb}
\widehat{U}=\widehat{U}_0(1-i\widehat{T}),
\end{equation}
where~$\widehat{U}_0$ is the evolution operator with only~$\widehat{H}_0$ (ideal evolution) and~$\widehat{T}$ is the noise correction. We define the fidelity of a noisy gate with a specific starting state~$\ket{\psi_0}$ as
\begin{equation}
F=\left|\bra{\psi_0}\widehat{U}_0^\dagger\widehat{U}\ket{\psi_0}\right|^2.
\end{equation}
The gate error is then~\cite{Anikin2025Fast}
\begin{equation}
\label{eq:error_specific}
\epsilon(\psi_0)=1-F=\bra{\psi_0}\widehat{T}^\dagger\widehat{T}\ket{\psi_0}-\left|\bra{\psi_0}\widehat{T}\ket{\psi_0}\right|^2.
\end{equation}
The noise correction can be approximated by the first-order term of the Dyson series:
\begin{equation}
\label{eq:T_dyson}
\widehat{T}\approx\int\limits_0^{\tau}\widehat{U}_0^\dagger(t')\widehat{V}(t')\widehat{U}_0(t')\,dt',
\end{equation}
where~$\tau$ is the gate time.

We are interested in evaluating the mean error over all starting states. For the purposes of this work, we consider all starting states to be separable into the pure qubit part and the motional part, with the latter assumed to be a diagonal density matrix (i.e. a thermal distribution):
\begin{equation}
\rho = \ketbra{\psi}{\psi}\otimes\sum\limits_{n}p_n\ketbra{n}{n},
\end{equation}
where~$\ket{\psi}$ is a random two-qubit state, $\ket{n}$ is the multi-mode Fock basis state, and~$p_n$ is the probability of being in this state ($n$ is thought of as a multi-index running over all modes and their motional numbers). It can be shown (see Appendix~\ref{app:mean_error}) that the mean error in this case is calculated in a similar way to~\eqref{eq:error_specific}:
\begin{equation}
\label{eq:error_average}
\epsilon = \dfrac{1}{5}\sum\limits_n p_n \left[\Tr(\widehat{T}^{\dagger}\widehat{T})\right]_{nn} - \dfrac{1}{20}\sum\limits_{mn} p_m \left|(\Tr\widehat{T})_{nm}\right|^2,
\end{equation}
where~$\widehat{T}_{nm} = \braket*[3]{n}{\widehat{T}}{m}$, and the trace is taken over the two-qubit space.

The ideal evolution operator can be approximated by neglecting the carrier terms in both Hamiltonians~\eqref{eq:MS_hamiltonian_sensitive} and~\eqref{eq:MS_hamiltonian_insensitive}, which are the first sums in each expression (without the motional operators). This approximation is valid in the limit~$\Omega_j\ll\mu$~\cite{kirchmair_deterministic_2009,roos_ion_2008}, which holds when the gate time is much longer than the period of motional oscillations of ions in the trap. The noiseless evolution operator is therefore
\begin{equation}
\label{eq:noiseless_evolution_operator}
\widehat{U}_0(t)=e^{i\chi(t)\sigma_{\varphi}^{(1)}\sigma_{\varphi}^{(2)}}\prod\limits_l\mathcal{D}_l\left(\sum\limits_j\alpha_{lj}(t)\sigma_{\varphi}^{(j)}\right),
\end{equation}
where~$\chi(t),\alpha_{lj}(t)$ are the entanglement phase and motional mode displacements respectively~\cite{kirchmair_deterministic_2009,roos_ion_2008}, $\varphi$ is either~$\phi$ or~$\phi_\perp$, depending on the beam geometry, and~$\mathcal{D}_l(\alpha)$ is the displacement operator acting in the phase space of the~$l^{\mathrm{th}}$ mode. A more convenient expression for this operator is obtained is the eigenbasis of~$\sigma_{\varphi}^{(j)}$. The (single-qubit) eigenvalues and eigenstates are denoted~$s_j$ and~$\ket{s_j}$, respectively, where~$s_j\in\lbrace+1,-1\rbrace$. In this basis, the evolution operator takes the form
\begin{equation}
\widehat{U}_0(t)=\sum\limits_{s}\ketbra{s}{s}\,e^{i\chi(t)s_1s_2}\prod\limits_l\mathcal{D}_l\left(\sum\limits_j\alpha_{lj}(t)s_j\right),
\end{equation}
where~$\ket{s}=\ket{s_1s_2}$ is the two-qubit eigenstate. In this representation, we arrive to the following expression for~$\widehat{T}$, using~\eqref{eq:T_dyson}:
\begin{equation}
\label{eq:T_approx}
\widehat{T}\approx\sum\limits_{ss'}\ketbra{s}{s'}\int\limits_0^{\tau}e^{-i\chi'(s_1s_2 - s_1's_2')}\prod\limits_l\mathcal{D}_l\left(-\sum\limits_j\alpha_{lj}'s_j\right)\widehat{V}'_{ss'}\prod\limits_l\mathcal{D}_l\left(\sum\limits_j\alpha_{lj}'s_j'\right)\,dt',
\end{equation}
where~$\alpha_{lj}'=\alpha_{lj}(t'),\chi'=\chi(t'),\widehat{V}'=\widehat{V}(t')$. Since the trace is invariant with respect to basis changes, it is straightforward to
\textcolor{black}{use~\eqref{eq:T_approx} to}
obtain the terms in~\eqref{eq:error_average}:
\begin{gather}
\label{eq:Tr_T}
\Tr\widehat{T}\approx\sum\limits_{s}\int\limits_0^{\tau}\prod\limits_l\mathcal{D}_l\left(-\sum\limits_j\alpha_{lj}'s_j\right)\widehat{V}'_{ss}\prod\limits_l\mathcal{D}_l\left(\sum\limits_j\alpha_{lj}'s_j\right)\,dt',\\\allowdisplaybreaks
\begin{multlined}
\label{eq:Tr_T+T}
\Tr(\widehat{T}^{\dagger}\widehat{T})\approx\sum\limits_{ss'}\iint e^{i(\chi'-\chi'')(s_1s_2 - s_1's_2')-i\varepsilon(t',t'')}\\
\times\prod\limits_l\mathcal{D}_l\left(-\sum\limits_j\alpha_{lj}'s_j'\right)\widehat{V}'_{s's}\prod\limits_l\mathcal{D}_l\left(\sum\limits_j(\alpha_{lj}' - \alpha_{lj}'')s_j\right)\widehat{V}''_{ss'}\prod\limits_l\mathcal{D}_l\left(\sum\limits_j\alpha_{lj}''s_j'\right)\,dt'dt'',
\end{multlined}
\end{gather}
where
\begin{equation}
\varepsilon(t',t'') = \sum\limits_{jj'}s_js_{j'}\sum\limits_l\Im(\alpha_{lj}'\alpha_{lj'}''^*)
\end{equation}
appeared from the identity
\begin{equation}
\mathcal{D}(\alpha)\mathcal{D}(\beta)=e^{i\Im(\alpha\beta^*)}\mathcal{D}(\alpha+\beta).
\end{equation}
The limits of double integration are omitted in~\eqref{eq:Tr_T+T} and henceforth for clarity. We use~\eqref{eq:Tr_T} and~\eqref{eq:Tr_T+T} to calculate the error in each spectral range for each laser configuration, with respective noise Hamiltonians~$\widehat{V}$.

\subsection{Fast noise}
\label{sec:fast_noise}
This section is devoted to the analysis of the fast noise terms in~\eqref{eq:noise_hamiltonian_sensitive} and~\eqref{eq:noise_hamiltonian_insensitive}. For both configurations, the noise terms commute with the displacement operators~$\mathcal{D}_l$, since they do not contain any motional parts. Taking this into account, the expressions~\eqref{eq:Tr_T} and~\eqref{eq:Tr_T+T} will be simplified as
\begin{gather}
\label{eq:Tr_T_commute}
\Tr\widehat{T}\approx\sum\limits_{s}\int\limits_0^{\tau}\widehat{V}'_{ss}\,dt',\\\allowdisplaybreaks
\label{eq:Tr_T+T_commute}
\begin{multlined}
\Tr(\widehat{T}^{\dagger}\widehat{T})\approx\sum\limits_{ss'}\iint \exp\left[i(\chi'-\chi'')(s_1s_2 - s_1's_2')+i(\varepsilon'-\varepsilon'')(s_1s_2'-s_1's_2)-i\theta_{ss'}(t',t'')\right]\\
\times\widehat{V}'_{s's}\widehat{V}''_{ss'}\prod\limits_l\mathcal{D}_l\left(\sum\limits_j(\alpha_{lj}'-\alpha_{lj}'')(s_j - s_j')\right)\,dt'dt'',
\end{multlined}
\end{gather}
where
\begin{equation}
\begin{gathered}
\varepsilon'= \Im\sum\limits_l \alpha'_{l1}\alpha'^*_{l2},\quad \varepsilon''= \Im\sum\limits_l \alpha''_{l1}\alpha''^*_{l2},\\
\theta_{ss'}(t',t'') = \sum\limits_{jj'}(s_j - s'_j)(s_{j'} - s'_{j'})\Im\sum\limits_l\alpha'_{lj}\alpha''^*_{lj'}.
\end{gathered}
\end{equation}
We now proceed to calculating the gate error in each configuration, using~\eqref{eq:Tr_T_commute} and~\eqref{eq:Tr_T+T_commute}.
\subsubsection{Phase-sensitive configuration}
In this case we get
\begin{equation}
\widehat{V}^{(\mathrm{S})}_{ss'} = -\delta\phi(t)\delta_{ss'}\sum\limits_j \Omega_j s_j\cos(\mu t + \phi_M^{(j)}).
\end{equation}
Note that
\begin{equation}
\sum\limits_s\widehat{V}^{(\mathrm{S})}_{ss} = -\delta\phi(t)\sum\limits_j \Omega_j \cos(\mu t + \phi_M^{(j)})\sum\limits_s s_j = 0,
\end{equation}
therefore,~$\Tr\widehat{T} = 0$. For~$\Tr(\widehat{T}^{\dagger}\widehat{T})$, we obtain
\begin{equation}
\Tr(\widehat{T}^{\dagger}\widehat{T}) = \sum\limits_s\iint\delta\phi(t')\delta\phi(t'')\sum\limits_{jj'}s_js_{j'}\Omega'_j\Omega''_{j'} \cos(\mu t' + \phi_M^{(j)}) \cos(\mu t'' + \phi_M^{(j')})\,dt'dt''.
\end{equation}
Firstly, we make use of the fact that
\begin{equation}
\label{eq:sum_s_js_j'}
\sum\limits_s s_js_{j'} = 4\delta_{jj'}.
\end{equation}
Secondly, assuming the noise~$\delta\phi(t)$ is stationary, we use the Wiener–Khinchin theorem:
\begin{equation}
\label{eq:wiener_khinchin}
\left\langle\delta\phi(t')\delta\phi(t'')\right\rangle = \int\limits_0^{+\infty}S^{(1)}_{\phi}(f)\cos(2\pi f(t'-t''))\,df,
\end{equation}
where~$S^{(1)}_{\phi}(f)$ is the one-sided phase noise PSD. Substituting~\eqref{eq:sum_s_js_j'} and~\eqref{eq:wiener_khinchin} into~\eqref{eq:error_average}, we obtain the expression for the error, averaged over all noise trajectories:
\begin{equation}
\label{eq:error_sensitive_convolution}
\epsilon^{(\mathrm{S})}_{\mathrm{fast}} = \int\limits_0^{+\infty}S^{(1)}_{\phi}(f)\mathcal{F}^{(\mathrm{S})}_{\mathrm{fast}}(f)\,df,
\end{equation}
where the \emph{filter function}~$\mathcal{F}^{(\mathrm{S})}(f)$ \textcolor{black}{(in Hz/Hz)} is
\begin{equation}
\label{eq:F_fast_sensitive_first}
\mathcal{F}^{(\mathrm{S})}_{\mathrm{fast}}(f) = \dfrac{4}{5}\sum\limits_j \left|\int\limits_0^{\tau}e^{2\pi ift'}\Omega'_j\cos(\mu t' + \phi_M^{(j)})\,dt'\right|^2.
\end{equation}
For the simplest case of constant laser power ($\Omega_j = \mathrm{const}$) the time integral can be calculated analytically. From the form of~$\mathcal{F}^{(\mathrm{S})}(f)$ it is clear that the filter function will be peaked (in Fourier domain) at~$\mu/2\pi$. Around this peak, assuming~$\mu\tau\gg 1$ (this is the same assumption that was made when the carrier term was neglected earlier), and a uniform laser power across both ions:~$\Omega_j=\Omega=\mathrm{const}$, we arrive at a simple formula for the filter function:
\begin{equation}
\label{eq:F_fast_sensitive}
\mathcal{F}^{(\mathrm{S})}_{\mathrm{fast}}(f) = \dfrac{2(\Omega\tau)^2}{5}\sinc^2\left(\dfrac{(2\pi f-\mu)\tau}{2}\right).
\end{equation}
\textcolor{black}{This result can be generalized to make its physical meaning clearer. Many state-of-the-art trapped-ion setups use some form of pulse shaping of the driving field, that is, a certain modulation of the pulse which helps reduce errors~\cite{blumel_power-optimal_2021, weber_robust_2024, hughes2025trapped, debnath_demonstration_2016, zhu_arbitraryspeed_2006, milne_phasemodulated_2020, choi_optimal_2014, shapira_theory_2020}. A general shaped pulse will have its driving function changed in the following way:
\begin{equation}
\Omega_j(t)\cos(\mu t+\phi_M^{(j)})\rightarrow g_j(t),
\end{equation}
where~$g_j(t)$ is the pulse function, that is both amplitude and frequency modulated. In this case, we can see that the filter function~\eqref{eq:F_fast_sensitive_first} is, up to a constant factor and the sum over all ions, the squared $\tau$-windowed Fourier transform of the pulse function. For a constant amplitude (square) pulse, the Fourier transform is known to be the sinc function centered at~$\mu/(2\pi)$, which equation~\eqref{eq:F_fast_sensitive} shows. In a more complicated case of an amplitude modulated pulse, such as in~\cite{debnath_demonstration_2016} or~\cite{zhu_arbitraryspeed_2006}, the filter function will be centered around the "carrier frequency" of the pulse~$\mu/(2\pi)$, and the sideband Fourier components of the pulse envelope will be placed at frequencies~$\sim 1/\tau$ from the carrier. Those components will be additionally "smeared out" by approximately~$1/\tau$ as well due to the convolution with the sinc function, which is the Fourier transform of the square pulse, as in the constant amplitude case. For an amplitude \emph{and} frequency modulated pulse, we can expect its most prominent Fourier components to be around one or several normal mode frequencies of the ionic chain~\cite{blumel_power-optimal_2021}, therefore, the filter function will still be localized around the motional spectrum and "smeared" by~$\sim 1/\tau$.
}

\subsubsection{Phase-insensitive configuration}
\label{sec:fast_insensitive}
For the phase-insensitive configuration, we calculate the matrix elements of~$\sigma_{\phi}^{(j)}$ between two eigenstates of~$\sigma_{\phi_\perp}^{(1)}\sigma_{\phi_\perp}^{(2)}$:
\begin{equation}
\label{eq:sigma_matrix_element}
(\sigma_{\phi}^{(j)})_{ss'} = is'_j\,\delta_{s_k,s'_k}\delta_{s_j,-s'_j}=-is_j\,\delta_{s_k,s'_k}\delta_{s_j,-s'_j},
\end{equation}
where~$k$ denotes the index of the other ion (not~$j$). This means~$\widehat{V}_{ss} = 0$, so~$\Tr\widehat{T} = 0$. Additionally, we can simplify~\eqref{eq:Tr_T+T_commute} by using
\begin{multline}
\widehat{V}'_{s's}\widehat{V}''_{ss'} \propto \sum\limits_{jj'}\Omega'_j\Omega''_{j'} \cos(\mu t' + \phi_M^{(j)}) \cos(\mu t'' + \phi_M^{(j')})s_js_{j'}\delta_{s_k,s'_k}\delta_{s_j,-s'_j}\delta_{s_{k'},s'_{k'}}\delta_{s_{j'},-s'_{j'}}\\
=\sum\limits_{jj'}\Omega'_j\Omega''_{j'} \cos(\mu t' + \phi_M^{(j)}) \cos(\mu t'' + \phi_M^{(j')})s_js_{j'}\delta_{jj'}\delta_{s_k,s'_k}\delta_{s_j,-s'_j}\\
=\sum\limits_{j}\Omega'_j\Omega''_{j} \cos(\mu t' + \phi_M^{(j)}) \cos(\mu t'' + \phi_M^{(j)})\delta_{s_k,s'_k}\delta_{s_j,-s'_j}.
\end{multline}
This expression means that fixing~$s$ and~$j$ fixes~$s'$, so the sum over~$s'$ in~\eqref{eq:Tr_T+T_commute} is redundant. Furthermore, we can use the expression above to obtain (for fixed~$s,j$)
\begin{equation}
\begin{gathered}
s_1s_2 - s_1's_2' = 2s_1s_2,\quad s_1s_2'-s_1's_2=2s_1s_2',\\
\theta_{ss'}=4\Im\sum\limits_l\alpha'_{lj}\alpha''^*_{lj},\\
\sum\limits_j(\alpha_{lj}'-\alpha_{lj}'')(s_j - s_j')=2s_j(\alpha_{lj}'-\alpha_{lj}'').
\end{gathered}
\end{equation}
Substituting into~\eqref{eq:Tr_T+T_commute}, we arrive at
\begin{multline}
\Tr(\widehat{T}^{\dagger}\widehat{T})\approx\sum\limits_{s,j}\iint \exp\left[2is_1s_2(\chi'-\chi'')+2is_1s_2'(\varepsilon'-\varepsilon'')-4i\Im\sum\limits_l\alpha'_{lj}\alpha''^*_{lj}\right]\\
\times\delta\phi(t')\delta\phi(t'')\Omega'_j\Omega''_{j} \cos(\mu t' + \phi_M^{(j)}) \cos(\mu t'' + \phi_M^{(j)})\prod\limits_l\mathcal{D}_l\left(2s_j(\alpha_{lj}'-\alpha_{lj}'')\right)\,dt'dt'',
\end{multline}
where~$s'$ is assumed to be a function of~$s,j$, defined by
\textcolor{black}{
\begin{equation}
s'_j(s,j) = -s_j,\quad s'_k(s,j) = s_k.
\end{equation}
The arguments~$(s,j)$ are omitted henceforth for clarity.} To calculate the sum over the motional states in~\eqref{eq:error_average}, we need to evaluate
\begin{equation}
\braket*[3]{n}{\prod\limits_l\mathcal{D}_l\left(2s_j(\alpha_{lj}'-\alpha_{lj}'')\right)}{n}.
\end{equation}
The diagonal matrix element of the displacement operator is expressed in terms of the Laguerre polynomial~$L_n(x)$:
\begin{equation}
\braket[3]{n_l}{\mathcal{D}_l(\alpha))}{n_l} = e^{-|\alpha|^2/2}L_{n_l}(|\alpha|^2).
\end{equation}
Assuming the motional density matrix to be a thermal state for all normal modes with the mean motional numbers~$\overline{n}_l$ in~$l$-th mode respectively, we get
\begin{equation}
p_n =\prod\limits_{l} p_{n_l}=\prod\limits_l\dfrac{1}{1+\overline{n}_l}\left(\dfrac{\overline{n}_l}{1+\overline{n}_l}\right)^{n_l}.
\end{equation}
The sum over all~$n_l$ can be calculated using the generating function for~$L_n$~\cite{abramowitz_handbook_1965}:
\begin{equation}
\sum\limits_{n=0}^{\infty}L_n(x)t^n = \dfrac{e^{-xt/(1-t)}}{1-t}.
\end{equation}
Substituting~$t=\overline{n}_l/(1+\overline{n}_l),\ x=|\alpha|^2$, we get
\begin{equation}
\sum\limits_{n_l=0}^{\infty}p_{n_l}\braket[3]{n_l}{\mathcal{D}_l(\alpha))}{n_l}=\exp\left[-\left(\overline{n}_l+\dfrac{1}{2}\right)|\alpha|^2\right].
\end{equation}
Finally, using~\eqref{eq:wiener_khinchin}, we arrive at an expression similar to~\eqref{eq:error_sensitive_convolution}, but with a different filter function
\begin{multline}
\label{eq:F_fast_insensitive}
\mathcal{F}^{(\mathrm{I})}_{\mathrm{fast}}(f)=\dfrac{1}{5}\sum\limits_{s,j}\iint \exp\left[2is_1s_2(\chi'-\chi'')
+2is_1s_2'(\varepsilon'-\varepsilon'')-4i\Im\sum\limits_l\alpha'_{lj}\alpha''^*_{lj}\right.\\
\left. - 4\sum\limits_l\left(\overline{n}_l+\dfrac{1}{2}\right)|\alpha_{lj}'-\alpha_{lj}''|^2\right]\cos(2\pi f(t'-t''))\\
\times\Omega'_j\Omega''_{j} \cos(\mu t' + \phi_M^{(j)}) \cos(\mu t'' + \phi_M^{(j)})\,dt'dt''.
\end{multline}
There is no closed-form expression for this function, since it now depends on the dynamics of the displacements~$\alpha_{lj}(t)$ and the entangling phase~$\chi(t)$, but some qualitative analysis can still be performed
\textcolor{black}{using the same reasoning as in the phase-sensitive case. Taking into account that the filter function is real, the integral can be represented as
\begin{multline}
\iint h(t',t'')\cos(2\pi f(t'-t''))g_j(t')g_j(t'')\,dt'dt'' \\
= \Re\iint h(t',t'')e^{2\pi if(t'-t'')}g_j(t')g_j(t'')\,dt'dt'',
\end{multline}
where
\begin{multline}
h(t',t'') = \cos\ab[2s_1s_2(\chi'-\chi'')
+2s_1s_2'(\varepsilon'-\varepsilon'')-4\Im\sum\limits_l\alpha'_{lj}\alpha''^*_{lj}]\\
\times\exp\ab[- 4\sum\limits_l\left(\overline{n}_l+\dfrac{1}{2}\right)|\alpha_{lj}'-\alpha_{lj}''|^2],
\end{multline}
and the general pulse function~$g_j(t)$ is used. The function~$h$ is continuous and symmetric: $h(t',t'')=h(t'',t')$, therefore, it can be decomposed inside the area of integration using the Hilbert-Schmidt theorem as follows:
\begin{equation}
h(t',t'')=\sum\limits_n c_n\varphi_n(t')\varphi_n(t''),
\end{equation}
with both~$c_n$ and~$\varphi_n(t)$ being real. Then the integral will become
\begin{equation}
\sum\limits_n c_n\Re\int\limits_0^{\tau}e^{2\pi ift'}\varphi_n(t')g_j(t')\,dt'\int\limits_0^{\tau}e^{-2\pi ift''}\varphi_n(t'')g_j(t'')\,dt''=\sum\limits_n c_n\ab|\int\limits_0^{\tau}e^{2\pi ift}\varphi_n(t)g_j(t)\,dt|^2.
\end{equation}
The final expression has the same form of the squared windowed Fourier transform as in the phase-sensitive geometry. Since the functions~$\varphi_n(t)$ depend on $\alpha_{lj}(t),\chi(t)$, which evolve on the time scale of the total gate duration~$\tau$, the Fourier components of~$\varphi_n(t)$ will span a bandwidth of approximately~$1/\tau$, and the "carrier" frequency of the transform will be close to the motional spectrum for the same reason it was in the phase-sensitive case. Therefore, we expect both the phase-sensitive and phase-insensitive fast noise filter functions to be located around the detuning frequency and have widths of roughly the inverse gate time.}

\subsection{Slow noise}
\label{sec:slow_noise}
We begin this section by comparing the expressions for the Hamiltonians and the slow noise terms in both geometries. For the phase-sensitive geometry:
\begin{equation}
\label{eq:slow_sensitive_H_V}
\begin{gathered}
\widehat{H}_{\mathrm{MS}}^{(\mathrm{S})}(t)\approx
-\sum\limits_{l,j} \eta_{lj}\,\Omega_j \sigma_{\phi}^{(j)}\cos(\mu t+\phi_M^{(j)})
(\hat{a}_le^{-i\omega_l t} + \hat{a}^{\dagger}_le^{i\omega_l t}),
\\\allowdisplaybreaks
\widehat{V}^{(\mathrm{S})}(t)=-\delta\phi(t)\sum\limits_{l,j} \eta_{lj}\,\Omega_j \sigma_{\phi_\perp}^{(j)}\cos(\mu t+\phi_M^{(j)})
(\hat{a}_le^{-i\omega_l t} + \hat{a}^{\dagger}_le^{i\omega_l t}).
\end{gathered}
\end{equation}
For the phase-insensitive geometry:
\begin{equation}
\begin{gathered}
\widehat{H}_{\mathrm{MS}}^{(\mathrm{I})}(t)\approx\sum\limits_{l,j} \eta_{lj}\,\Omega_j \sigma_{\phi_\perp}^{(j)}\sin(\mu t+\phi_M^{(j)})
(\hat{a}_le^{-i\omega_l t} + \hat{a}^{\dagger}_le^{i\omega_l t}),
\\\allowdisplaybreaks
\widehat{V}^{(\mathrm{I})}(t)=-\delta\phi(t)\sum\limits_{l,j} \eta_{lj}\,\Omega_j \sigma_{\phi}^{(j)}\sin(\mu t+\phi_M^{(j)})
(\hat{a}_le^{-i\omega_l t} + \hat{a}^{\dagger}_le^{i\omega_l t}).
\end{gathered}
\end{equation}
It is evident that up to an inconsequential sign change, as well as up to a redefinition of the phases~$\phi_M^{(j)},\phi_S^{(j)}$, the expressions for both configurations are identical. Therefore, we may consider only one of them in our analysis and then redefine the phases for the other one. Henceforth, we take the expressions~\eqref{eq:slow_sensitive_H_V} as primary and omit the index~$(\mathrm{S})$.

The next step in our analysis is to simplify the problem by applying a unitary transformation of the following form:
\begin{equation}
\label{eq:U_R}
\widehat{U}_R(t) = \exp\left(-i\dfrac{\delta\phi(t)}{2}\sum\limits_j\sigma_z^{(j)}\right).
\end{equation}
In the new reference frame defined by~$\widehat{U}_R$, the full effective Hamiltonian will be
\begin{equation}
\label{eq:Hamiltonian_rotated}
\widehat{H}_R = \widehat{U}_R\widehat{H}_{\mathrm{MS}}\widehat{U}_R^{\dagger} + \widehat{U}_R\widehat{V}\widehat{U}_R^{\dagger} +i\dfrac{\partial\widehat{U}_R}{\partial t}\widehat{U}_R^{\dagger}.
\end{equation}
To see why this transformation simplifies the problem, we write~\eqref{eq:slow_sensitive_H_V} as follows:
\begin{equation}
\begin{gathered}
\widehat{H}_{\mathrm{MS}}(t) = \sum\limits_j \sigma_{\phi}^{(j)}\widehat{W}_j(t),\\
\widehat{V}(t) = \delta\phi(t)\sum\limits_j \sigma_{\phi_\perp}^{(j)}\widehat{W}_j(t),
\end{gathered}
\end{equation}
where~$\widehat{W}_j(t)$ is the motional part of the operator, identical in both terms. Using the commutation relations between Pauli matrices, we can show that
\begin{equation}
\begin{gathered}
\widehat{U}_R\sigma_{\phi}^{(j)}\widehat{U}_R^{\dagger} = \sigma_{\phi}^{(j)}\cos\delta\phi - \sigma_{\phi_\perp}^{(j)}\sin\delta\phi\approx\sigma_{\phi}^{(j)} - \sigma_{\phi_\perp}^{(j)}\delta\phi,\\
\widehat{U}_R\sigma_{\phi_\perp}^{(j)}\widehat{U}_R^{\dagger} = \sigma_{\phi_\perp}^{(j)}\cos\delta\phi + \sigma_{\phi}^{(j)}\sin\delta\phi\approx\sigma_{\phi_\perp}^{(j)} + \sigma_{\phi}^{(j)}\delta\phi
\end{gathered}
\end{equation}
up to linear terms in~$\delta\phi$. From this, neglecting~$(\delta\phi)^2$, we obtain the first two terms in~\eqref{eq:Hamiltonian_rotated}:
\begin{multline}
\widehat{U}_R\widehat{H}_{\mathrm{MS}}\widehat{U}_R^{\dagger} + \widehat{U}_R\widehat{V}\widehat{U}_R^{\dagger} \approx \sum\limits_j (\sigma_{\phi}^{(j)} - \sigma_{\phi_\perp}^{(j)}\delta\phi(t))\widehat{W}_j(t)
+ \delta\phi(t)\sum\limits_j (\sigma_{\phi_\perp}^{(j)} + \sigma_{\phi}^{(j)}\delta\phi(t))\widehat{W}_j(t)\\
\approx\sum\limits_j \sigma_{\phi}^{(j)}\widehat{W}_j(t) = \widehat{H}_{\mathrm{MS}}(t).
\end{multline}
The last term in~\eqref{eq:Hamiltonian_rotated} is
\begin{equation}
\label{eq:noise_rotating_frame}
i\dfrac{\partial\widehat{U}_R}{\partial t}\widehat{U}_R^{\dagger} = \dfrac{\delta\dot{\phi}(t)}{2}\sum\limits_j\sigma_z^{(j)}=\dfrac{\delta\omega(t)}{2}\sum\limits_j\sigma_z^{(j)},
\end{equation}
where~$\delta\omega(t) = \delta\dot{\phi}(t)$ is the frequency deviation. The full effective Hamiltonian consists of the laboratory frame unperturbed Hamiltonian~$\widehat{H}_{\mathrm{MS}}$ and an additional term~\eqref{eq:noise_rotating_frame}, which can be treated as the noise term in the new rotating frame.

Note that the noise in this frame does not contain any motional operators; therefore, the same argument that was made for the fast noise is valid here, and~\eqref{eq:Tr_T_commute} and~\eqref{eq:Tr_T+T_commute} apply. Furthermore, the matrix elements of~$\sigma_z^{(j)}$ between two eigenstates of~$\sigma_{\phi}^{(1)}\sigma_{\phi}^{(2)}$ are
\begin{equation}
(\sigma_z^{(j)})_{ss'} = -\delta_{s_k, s_k'}\delta_{s_j, -s_j'},
\end{equation}
which is identical to~\eqref{eq:sigma_matrix_element} up to a constant factor of magnitude 1. Thus, all calculations made in section~\ref{sec:fast_insensitive} will be the same here, up to the change
\begin{equation}
\delta\phi(t)\Omega_j\cos(\mu t+\phi_M^{(j)})\rightarrow\dfrac{\delta\omega(t)}{2}.
\end{equation}
The gate error will be of the form~\eqref{eq:error_sensitive_convolution}, but with frequency noise PSD~$S_{\omega}(f)$ instead of~$S_{\phi}(f)$:
\begin{equation}
\label{eq:error_slow_convolution}
\epsilon_{\mathrm{slow}} = \int\limits_0^{+\infty}S^{(1)}_{\omega}(f)\mathcal{F}_{\mathrm{slow}}(f)\,df,
\end{equation}
where the filter function~$\mathcal{F}_{\mathrm{slow}}(f)$ \textcolor{black}{(in (rad/s)$^{-2}$, note the different units compared to $\mathcal{F}_{\mathrm{fast}}(f)$)} is obtained from~\eqref{eq:F_fast_insensitive} with the aforementioned variable change:
\begin{multline}
\label{eq:F_slow}
\mathcal{F}_{\mathrm{slow}}^{\mathrm{(rot)}}(f)=\dfrac{1}{20}\sum\limits_{s,j}\iint \exp\left[2is_1s_2(\chi'-\chi'')
+2is_1s_2'(\varepsilon'-\varepsilon'')-4i\Im\sum\limits_l\alpha'_{lj}\alpha''^*_{lj}\right.\\
\left. - 4\sum\limits_l\left(\overline{n}_l+\dfrac{1}{2}\right)|\alpha_{lj}'-\alpha_{lj}''|^2\right]\cos(2\pi f(t'-t''))\,dt'dt''.
\end{multline}

This expression is valid in the rotating frame (hence the superscript "(rot)"), where the ideal gate evolution operator is assumed to be identical to the noiseless evolution operator~$\widehat{U}_0$ from~\eqref{eq:noiseless_evolution_operator}. However, in the lab frame, the noiseless evolution operator will be
\begin{equation}
\widehat{U}'_0 = \widehat{U}^{\dagger}_R(\tau)\widehat{U}_0
\end{equation}
and the ideal evolution operator is still~$\widehat{U}_0$. This leads to a modified expression for the fidelity, because the unitary transformation~\eqref{eq:U_R} is itself dependent on the same phase noise. If the noise correction induced by~\eqref{eq:noise_rotating_frame} is~$\widehat{T}$, then the fidelity will be equal
\begin{equation}
\label{eq:fidelity_lab}
F=\left|\bra{\psi_0}\widehat{U}_0^\dagger\widehat{U}^{\dagger}_R(\tau)\widehat{U}_0(1-i\widehat{T})\ket{\psi_0}\right|^2.
\end{equation}
To find the filter function in this case, we calculate the factor $\widehat{U}_0^\dagger\widehat{U}^{\dagger}_R(\tau)\widehat{U}_0$.
Substituting~\eqref{eq:noiseless_evolution_operator} and~\eqref{eq:U_R} and expanding up to terms linear in~$\delta\phi$, we get
\begin{equation}
\widehat{U}_0^\dagger\widehat{U}^{\dagger}_R(\tau)\widehat{U}_0\approx1-i\widehat{Q},
\end{equation}
where
\begin{equation}
\widehat{Q} = \dfrac{\delta\phi(\tau)}{2}\sum\limits_{ss'}\ketbra{s}{s'}e^{-2is_1s_2\chi}\sum\limits_{j}\delta_{s_k, s_k'}\delta_{s_j, -s_j'}
\end{equation}
and~$\chi=\chi(\tau)$. The phase deviation~$\delta\phi$ can be expressed in terms of frequency deviation~$\delta\omega$ as
\begin{equation}
\delta\phi(\tau)=\int\limits_0^{\tau}\delta\omega(t')\,dt',
\end{equation}
so the rotation noise correction will be
\begin{equation}
\widehat{Q} = \dfrac{1}{2}\sum\limits_{ss'}\ketbra{s}{s'}\int\limits_0^{\tau}\delta\omega(t')e^{-2is_1s_2\chi}\sum\limits_{j}\delta_{s_k, s_k'}\delta_{s_j, -s_j'}\,dt'.
\end{equation}
Substituting into~\eqref{eq:fidelity_lab}, we obtain
\begin{equation}
\label{eq:F_T_Q}
F=\left|\bra{\psi_0}(1-i\widehat{Q})(1-i\widehat{T})\ket{\psi_0}\right|^2,
\end{equation}
which, up to linear terms in~$\delta\omega$ (or~$\delta\phi$), is
\begin{equation}
F=\left|\bra{\psi_0}(1-i(\widehat{Q} + \widehat{T}))\ket{\psi_0}\right|^2.
\end{equation}
Thus, the effective noise correction term will be~$\widehat{Q} + \widehat{T}$. For this noise correction, noting that~$\Tr\widehat{Q}=\Tr\widehat{T}=0$, the gate error~\eqref{eq:error_average} will be
\begin{equation}
\label{eq:error_slow_Q_T}
\epsilon_{\mathrm{slow}}^{\mathrm{(lab)}} = \dfrac{1}{5}\sum\limits_n p_n \left[\Tr(\widehat{T}^{\dagger}\widehat{T} + \widehat{Q}^{\dagger}\widehat{Q} + \widehat{T}^{\dagger}\widehat{Q} + \widehat{Q}^{\dagger}\widehat{T})\right]_{nn}.
\end{equation}
The~$\widehat{T}^{\dagger}\widehat{T}$ term will give the filter function~\eqref{eq:F_slow}. The remaining terms are calculated using the same techniques that were used for calculating the previous filter functions:
\begin{equation}
\begin{gathered}
\sum\limits_n p_n \left[\Tr(\widehat{Q}^{\dagger}\widehat{Q})\right]_{nn} = 2\iint\delta\omega(t')\delta\omega(t'')\,dt'dt'',\\\allowdisplaybreaks
\begin{multlined}
\sum\limits_n p_n \left[\Tr(\widehat{T}^{\dagger}\widehat{Q} + \widehat{Q}^{\dagger}\widehat{T})\right]_{nn} = -\dfrac{1}{2}\sum\limits_{s,j}\iint\delta\omega(t')\delta\omega(t'')\cos\left(2s_1s_2(\chi-\chi') - 2s_1s_2'\varepsilon'\right)\\\allowdisplaybreaks
\times\exp\left[- 4\sum\limits_l\left(\overline{n}_l+\dfrac{1}{2}\right)|\alpha_{lj}'|^2\right]\,dt'dt'',
\end{multlined}
\end{gathered}
\end{equation}
which will lead to the following filter function:
\begin{multline}
\label{eq:F_slow_lab}
\mathcal{F}_{\mathrm{slow}}^{\mathrm{(lab)}}(f)=\dfrac{1}{20}\iint\left\lbrace 8 + \sum\limits_{s,j} \left(\exp\left[2is_1s_2(\chi'-\chi'')\vphantom{\sum\limits_l}
+2is_1s_2'(\varepsilon'-\varepsilon'')-4i\Im\sum\limits_l\alpha'_{lj}\alpha''^*_{lj}\right.\right.\right.\\
\left. - 4\sum\limits_l\left(\overline{n}_l+\dfrac{1}{2}\right)|\alpha_{lj}'-\alpha_{lj}''|^2\right]
- 2\cos\left(2s_1s_2(\chi-\chi') - 2s_1s_2'\varepsilon'\right)\\
\left.\left.\times\exp\left[- 4\sum\limits_l\left(\overline{n}_l+\dfrac{1}{2}\right)|\alpha_{lj}'|^2\right]\right)\right\rbrace\cos(2\pi f(t'-t''))\,dt'dt''.
\end{multline}

\section{Results}
\label{sec:Results}
\subsection{Comparison of sensitive and insensitive configurations in terms of error induced by fast noise}
\begin{figure}[h]
    \centering
    \includegraphics[width=0.7\linewidth]{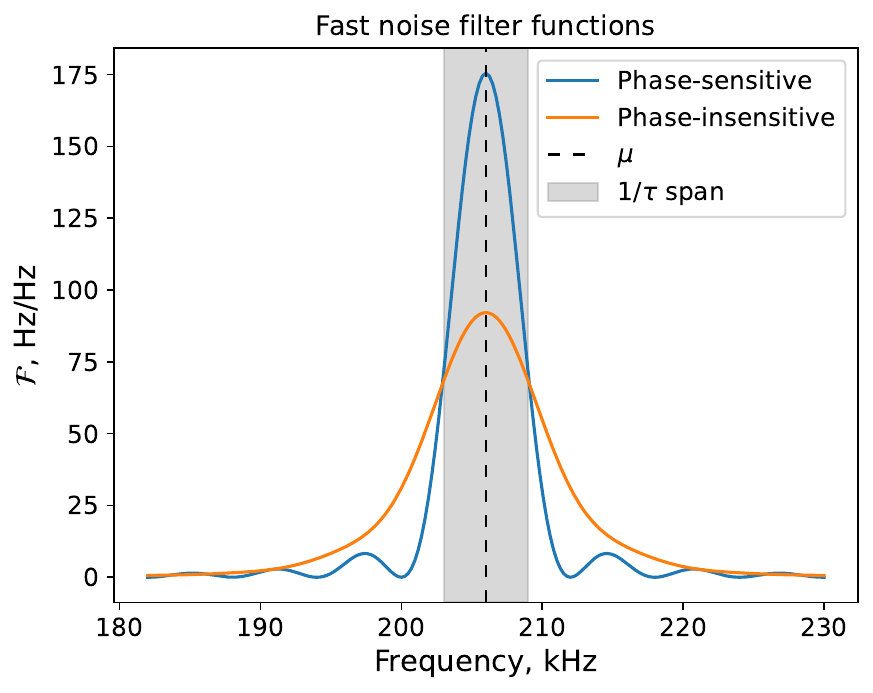}
    \caption{Filter functions for fast laser phase noise, both sensitive and insensitive configurations. The dashed line represents the detuning from the carrier transition~$\mu$. The width of the shaded region is~$1/\tau$, where~$\tau$ is the gate time. For other details of the calculation see main text}
    \label{fig:fast_psd}
\end{figure}

Based on the reasons listed earlier, the peak positions of the filter functions of the fast noise for both configurations should not differ much. This assumption is confirmed numerically, see fig.~\ref{fig:fast_psd}. The \textcolor{black}{filter functions} were calculated using the \textcolor{black}{same parameters that were used for the numerical simulations in~\cite{nakav_effect_2023}:
\begin{itemize}
    \item A single motional mode~$\omega_l=2\pi\cdot 200\ \text{kHz}$;
    \item Lamb-Dicke parameter for this mode~$\eta=0.15$;
    \item Rabi frequency (identical for both ions)~$\Omega=2\pi\cdot 20\ \text{kHz}$;
    \item Motional phases~$\phi_M^{(j)}=0$;
    \item Desired entanglement phase~$\chi_0=\pi/4$;
    \item Gate time~$\tau=\dfrac{2\pi}{\Omega\eta}\sqrt{\dfrac{\chi_0}{\pi}}\approx 167\ \mu\text{s}$~\cite{sorensen_entanglement_2000};
    \item Carrier detuning~$\mu=\omega_l+2\pi/\tau=2\pi\cdot 206\ \text{kHz}$.
\end{itemize}}
We see that the peak positions for both functions coincide with the detuning~$\mu$, and their widths are on the order of the inverse gate time. The filter function is slightly wider for the insensitive configuration, which can be attributed to the presence of other resonant frequencies that are close to~$\mu$ and effectively "smear out" the function \textcolor{black}{(see the previous discussion)}.

We can also obtain an estimate of the gate error for both cases, assuming the phase noise PSD changes slowly in frequency compared to the width of the filter function. In this case,~$S^{(1)}_\phi(f)$ can be moved out of the integral:
\begin{equation}
\label{eq:error_estimation_general}
\epsilon\approx S^{(1)}_\phi(f_0)\int\limits_{0}^{+\infty}\mathcal{F}(f)\,df,
\end{equation}
where~$f_0$ is the peak frequency of~$\mathcal{F}(f)$. For the phase-sensitive configuration, the integral of~\eqref{eq:F_fast_sensitive} can be calculated analytically in the limit~$\mu\tau\gg 1$:
\begin{equation}
\int\limits_0^{+\infty}\mathcal{F}^{(\mathrm{S})}_{\mathrm{fast}}(f)\,df \approx\int\limits_{-\infty}^{+\infty}\mathcal{F}^{(\mathrm{S})}_{\mathrm{fast}}(f)\,df = \dfrac{2}{5}\Omega^2\tau,
\end{equation}
therefore, in the phase-sensitive case:
\begin{equation}
\label{eq:error_fast_sensitive}
\epsilon^{(\mathrm{S})}_{\mathrm{fast}} = \dfrac{4}{5}S_\phi(\mu/2\pi)\Omega^2\tau,
\end{equation}
where~$S_\phi(f)=S^{(1)}_\phi(f)/2$ is the two-sided phase noise PSD. To calculate the integral of~\eqref{eq:F_fast_insensitive}, we change the order of integration (integrating over the frequency first) and use the identity
\begin{equation}
\label{eq:delta_trick}
\int\limits_0^{+\infty}\cos(2\pi f(t'-t''))\,df = \dfrac{1}{2}\delta(t'-t'').
\end{equation}
Assuming, as for the phase-sensitive case, a constant laser power that is uniform across the ions, we arrive at
\begin{equation}
\int\limits_0^{+\infty}\mathcal{F}^{(\mathrm{I})}_{\mathrm{fast}}(f)\,df = \dfrac{4}{5}\Omega^2\int\limits_0^{\tau}\cos^2(\mu t + \phi_M^{(j)})\,dt.
\end{equation}
Since we assume~$\mu\tau\gg 1$, the squared cosine in the integral can be approximated by~$1/2$, its average value. For the gate error, we obtain the expression
\begin{equation}
\label{eq:error_fast_insensitive}
\epsilon^{(\mathrm{I})}_{\mathrm{fast}} = \dfrac{4}{5}S_\phi(\mu/2\pi)\Omega^2\tau,
\end{equation}
which is exactly the same expression as the formula~\eqref{eq:error_fast_sensitive} for the error in the phase-sensitive configuration.
Thus, both laser beam geometries lead to the same gate error induced by fast noise.
\textcolor{black}{In the general modulation case, the approximate errors can be calculated in a similar way:
\begin{equation}
\epsilon^{(\mathrm{S})}_{\mathrm{fast}} = \epsilon^{(\mathrm{I})}_{\mathrm{fast}} \approx \dfrac{4}{5}S_\phi(\mu/2\pi)\int\limits_0^{\tau}\Omega^2(t)\,dt,
\end{equation}
where~$\Omega(t)$ is the envelope of the pulse function~$g(t)$ (assuming the same one for both ions). Therefore, equations~\eqref{eq:error_fast_sensitive} and~\eqref{eq:error_fast_insensitive} are applicable in this case as well, if~$\Omega$ is interpreted as the root mean square Rabi frequency of the pulse.}
\subsection{Comparison of the rotating and lab frames in terms of error induced by slow noise}
\begin{figure}[h]
    \centering
    \includegraphics[width=0.7\linewidth]{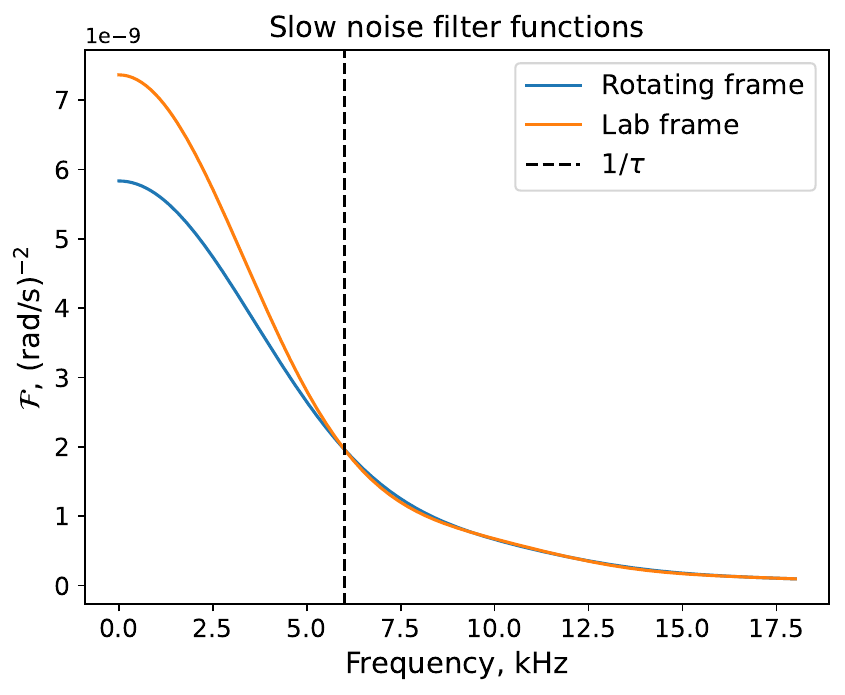}
    \caption{Slow laser frequency noise filter functions. The dashed line represents the value~$1/\tau$. \textcolor{black}{Note the different units compared to $\mathcal{F}_{\mathrm{fast}}(f)$}}
    \label{fig:slow_psd}
\end{figure}

We expect the slow noise filter function to be prominent at frequencies around~$\sim 1/\tau$ in both rotating and lab frames. Example functions are shown in fig.~\ref{fig:slow_psd}. The parameters were chosen identical to those used for the fast noise filter function in fig.~\ref{fig:fast_psd}.
The filter functions are indeed prominent at frequencies up to approximately the inverse gate time, which also turns out to be the value after which the functions are almost the same. Therefore, the difference between the rotating and lab frames in this case is significant only up to~$1/\tau$.

We can estimate the error in the case of slowly-varying frequency noise PSD for both frames in the same way as was done for fast noise: by moving the PSD out of the integral~\eqref{eq:error_slow_convolution}. \textcolor{black}{As was mentioned in the introduction,} the laser that performs the gate is usually stabilized by being locked to a frequency reference, such as a high-finesse Fabry-P\'erot cavity~\cite{kolachevsky2011low, Alnis2008Subhertz, schmid2019simple, Zalivako_2020}, making the low-frequency noise PSD flat (aka white frequency noise) up to approximately the servo bandwidth~\cite{Senko2022limits}.
We can then assume~$S^{(1)}_{\omega}(f)$ to be constant:
\begin{equation}
S^{(1)}_{\omega}(f) = h_0.
\end{equation}
The value~$h_0$ is connected to the linewidth of the laser~$\Delta\omega$ (FWHM) via the following identity~\cite{di2010simple}:
\begin{equation}
\label{eq:white_noise_linewidth}
h_0=2\Delta\omega.
\end{equation}
The integral~$\int\mathcal{F}(f)\,df$ can be calculated using~\eqref{eq:delta_trick}. Taking~\eqref{eq:white_noise_linewidth} into account, we obtain the errors as follows:
\begin{gather}
\label{eq:error_slow_rot_approx}
\epsilon_{\mathrm{slow}}^{\mathrm{(rot)}}\approx\dfrac{2}{5}\Delta\omega\tau,\\
\label{eq:error_slow_lab_approx}
\begin{multlined}
\epsilon_{\mathrm{slow}}^{\mathrm{(lab)}}\approx\dfrac{\Delta\omega}{20}\int\limits_0^{\tau}\left\lbrace 8 + \sum\limits_{s,j} \left(1\vphantom{\exp\left[- 4\sum\limits_l\left(\overline{n}_l+\dfrac{1}{2}\right)|\alpha_{lj}'|^2\right]}
- 2\cos\left(2s_1s_2(\chi-\chi') - 2s_1s_2'\varepsilon'\right)\right.\right.\\
\left.\left.\times\exp\left[- 4\sum\limits_l\left(\overline{n}_l+\dfrac{1}{2}\right)|\alpha_{lj}'|^2\right]\right)\right\rbrace\,dt'
\end{multlined}
\end{gather}
The expression~\eqref{eq:error_slow_lab_approx} depends on the gate parameters. However, it is straightforward to obtain the upper bound on the error, which is independent of these parameters:
\begin{equation}
\epsilon_{\mathrm{slow}}^{\mathrm{(lab)}}\leq\dfrac{\Delta\omega}{20}\int\limits_0^{\tau}\left\lbrace 8 + \sum\limits_{s,j}(1 + 2)\right\rbrace = \dfrac{8}{5}\Delta\omega\tau,
\end{equation}
which has the same order of magnitude as the error in the rotating frame~\eqref{eq:error_slow_rot_approx}.

\section{Discussion}
\label{sec:discussion}
\textcolor{black}{
It is important to emphasize the fact that the analysis that was done in the present work initially considered the intrinsic laser phase noise, that is, the initial noise that comes from the source of light and that is the same for all addressed ions and shifts the spin phase~$\phi_S$ (the noise that shifts the motional phase~$\phi_M$ was studied in~\cite{kang_designing_2023}). However, it is also applicable to all~$\phi_S$ noise that is common to all ions. For example, it can be the optical path fluctuations in the beam before it gets physically split into individual addressing beams, electronic noise in the generators that drive the electro-optical or acousto-optical modulators of the same beam or bias magnetic field fluctuations (assuming sufficient homogeneity of this field across the chain). The noise models and their respective PSDs will differ depending on the physical source of~$\phi_S$ noise, but the filter functions will remain the same and can be used to describe each such source.}

\textcolor{black}{Furthermore, if there is noise that influences only one of the~$\phi_S$ phases (like optical path fluctuations in one of the individual beams or electronic noise in an individual driving signal generator), our analysis can be adapted to that scenario as well. The noise Hamiltonians~\eqref{eq:noise_hamiltonian_sensitive} and~\eqref{eq:noise_hamiltonian_insensitive} will lose the ion index~$j$ in the sum, which will lead to all filter functions simply losing their respective summation index~$j$ as well. The fact that the filter function for both ions is simply the sum of individual filter functions for each ion leads to a following observation: if there are two different~$\phi_S$ noise functions~$\delta\phi_j(t)$, the~$\delta_{jj'}$ terms make it so that the average gate error will only depend on the individual noise autocorrelation functions~$\left\langle\delta\phi(t')\delta\phi(t'')\right\rangle$, even if the phase noise on the ions is partially correlated. Therefore, our analysis is directly extendable to any~$\phi_S$ noise of any nature as follows: the total gate error is the sum of two terms, one for each ion; each term is a convolution of the respective noise PSD (phase or frequency, for fast or slow noise, respectively) on that ion with the filter function of the gate for that ion. This is independent of any~$\phi_S$ fluctuation correlations between the ions.}

The influence of fast laser noise on the fidelity of two-qubit gates was studied before in~\cite{nakav_effect_2023}. In the referenced work the authors used numerical simulations to arrive at the following formula:
\begin{equation}
\label{eq:error_fast_RPSD}
1-F=T\cdot\mathrm{RPSD}(f_l),
\end{equation}
where~$T$ is the gate time, $f_l$ is the trap secular frequency, and~$\mathrm{RPSD}(f)$ is the so-called Rabi PSD, which is the laser field PSD, renormalized such that the area under the carrier peak is~$\Omega^2$. To compare this result with expressions~\eqref{eq:error_fast_sensitive} (or~\eqref{eq:error_fast_insensitive}) presented in this work, we make the following observations. First, the starting two-qubit state for the simulations in~\cite{nakav_effect_2023} was~$\ket{00}$. It can be shown that if the starting state is~$\ket{00}$ rather than a random one, the numerical factors~$4/5$ in~\eqref{eq:error_fast_sensitive} and~\eqref{eq:error_fast_insensitive} become unity.
Second, the detuning~$\mu$ is assumed to be close to the motional frequency, so~$S_\phi(f_l)\approx S_\phi(\mu/2\pi)$. Third, for small phase fluctuations, the electric field PSD of the laser at frequencies far from the carrier is proportional to the phase noise PSD~\cite{riehle_frequency_2006}, therefore, $S_\phi(f)\Omega^2\approx\mathrm{RPSD}(f)$, according to the definition of the latter. Combining all observations, we conclude that, under appropriate assumptions, the results obtained in this work agree with those obtained in~\cite{nakav_effect_2023}.

The fast noise parts in perturbations~\eqref{eq:noise_hamiltonian_sensitive} and~\eqref{eq:noise_hamiltonian_insensitive} come from the carrier term in the initial Hamiltonians~\eqref{eq:MS_hamiltonian_sensitive} and~\eqref{eq:MS_hamiltonian_insensitive}, respectively. However, this term was neglected in the noiseless evolution operator, while the noise terms themselves remained. Even though this does not seem justified at first, the errors introduced by the carrier term and the fast noise term are independent and therefore their calculation can be separated into two distinct problems. The argument is based on the same discussion that we made before when considering slow noise originating from uncorrelated sources. The fidelity in the presence of both systematic and stochastic errors can be represented in the form~\eqref{eq:F_T_Q}, where~$\widehat{Q}$ is the systematic error introduced by the carrier term. The cross terms in this case will be linear in the fluctuation, 
therefore, they cancel out in the mean error. This means that the total error with a systematic and stochastic source is the sum of two errors from each source separately. Note that the same is true for the case of several uncorrelated stochastic error sources, such as motional frequency drifts or magnetic field instability mentioned before. Furthermore, the systematic error operator~$\widehat{Q}$ can be any deterministic operator that represents the difference between the ideal noiseless evolution operator~\eqref{eq:noiseless_evolution_operator} and its desired action at the end of the gate~$e^{i\chi_0\sigma_{\varphi}^{(1)}\sigma_{\varphi}^{(2)}}$. Any deviation of the entanglement phase~$\chi(\tau)$ from~$\chi_0$ and the displacement parameters~$\alpha_{lj}(\tau)$ from zero can be incorporated into~$\widehat{Q}$ with their contribution calculated separately, while the stochastic error can be evaluated using the desired dynamics~$\chi(t), \alpha_{lj}(t)$, for which~$\chi(\tau)=\chi_0,\ \alpha_{lj}(\tau)=0$.

The difference between the rotating and the lab frame for the slow noise can be explained in the following way. In the rotating frame that is in phase with the laser (which~$\widehat{U}_R$ is an example of), the ideal MS gate Hamiltonian, as we have shown above, will be the same as the one in the non-rotating lab frame. Phase noise comes only as an additional~$\sigma_z$ term. The same is true for single-qubit gates. Therefore, if all gates in the quantum circuit are performed with the same laser source, the rotating frame will be the same for the entire evolution of the system, so it is possible to transition into this frame in the beginning of the circuit and move back into the lab frame once the whole circuit is finished. If the gate unitaries in the rotating frame are $\widehat{U}_1, \widehat{U}_2, \dots, \widehat{U}_N$, then the total evolution operator in the lab frame will be
\begin{equation}
\label{eq:evolution_circuit_lab}
\widehat{U}_{\mathrm{lab}} = \widehat{U}_R^{\dagger}(T_N)\widehat{U}_N\widehat{U}_{N-1}\dots\widehat{U}_1,
\end{equation}
where~$T_N$ is the total evolution time after~$N$ gates, and we assume~$\widehat{U}_R(0)$ is the identity. This means that in this case the additional error due to rotation can be ignored and accounted for separately at the end, so using the \textcolor{black}{rotating frame} filter function~\eqref{eq:F_slow} makes more sense. \textcolor{black}{Furthermore, the rotating frame is suitable for describing other~$\sigma_z$-type noises that are not correlated with the laser phase noise. A common example is magnetic field fluctuations, which affect the qubit frequency directly.}

\textcolor{black}{A different effect takes place when each gate is performed with a separate source uncorrelated with the rest. In this case, we use an} equivalent representation of~\eqref{eq:evolution_circuit_lab} \textcolor{black}{that} is the product of individual evolution operators in the lab frame:
\begin{multline}
\widehat{U}_{\mathrm{lab}} = \left[\widehat{U}_R^{\dagger}(T_N)\widehat{U}_N\widehat{U}_R(T_{N-1})\right]\\
\times\left[\widehat{U}_R^{\dagger}(T_{N-1})\widehat{U}_{N-1}\widehat{U}_R(T_{N-2})\right]\dots\left[\widehat{U}_R^{\dagger}(T_1)\widehat{U}_1\widehat{U}_R(0)\right]\\
=\widehat{U}^{\mathrm{(lab)}}_N\widehat{U}^{\mathrm{(lab)}}_{N-1}\dots\widehat{U}^{\mathrm{(lab)}}_1,
\end{multline}
where
\begin{equation}
\widehat{U}^{\mathrm{(lab)}}_n = \widehat{U}_R^{\dagger}(T_n)\widehat{U}_n\widehat{U}_R(T_{n-1}).
\end{equation}
In this picture, each individual operator gives its own evolution and error due to rotation. If each gate is performed by an independent source uncorrelated with the rest, then the rotation error must be accounted for in each gate separately, in which case the \textcolor{black}{lab frame filter} function~\eqref{eq:F_slow_lab} is the correct one. If there is only one gate in the circuit, both representations become identical, so the rotation error at the end of the gate cannot be separated from the error accumulated during the gate, and the function~\eqref{eq:F_slow_lab} will be the correct one in this case as well.

\textcolor{black}{Summing up, we list the use cases for both filter functions as follows:
\begin{itemize}
    \item Rotating frame
    \begin{enumerate}
        \item Evaluating the error of the MS gate in a multi-gate circuit implemented with the same laser source;
        \item Investigating other~$\sigma_z$-type noise sources, such as magnetic field fluctuations;
    \end{enumerate}
    \item Lab frame
    \begin{enumerate}
        \item Measuring the error of a single MS gate;
        \item Measuring the error of one MS gate in a multi-gate circuit implemented with different uncorrelated sources.
    \end{enumerate}    
\end{itemize}
Overall, despite these differences,}
the error estimates in both cases have been shown to be of the same order, so the universal approximation
\begin{equation}
\epsilon_{\mathrm{slow}}\sim\Delta\omega\tau
\end{equation}
is acceptable in most cases.

\textcolor{black}{The validity of the white-noise approximation that has been used for the simplified slow noise error estimate has an important caveat. Below very low frequencies (100~Hz and lower) the noise is dominated by~$1/f$ noise, as was pointed out in the introduction. Strictly speaking, the convolution integral will diverge. However, this integral only diverges for an infinite measurement time. For a finite measurement time~$T_0$, the error will be
\begin{equation}
\epsilon_T=\int\limits_{1/T_0}^{+\infty}S_{\omega}(f)\mathcal{F}(f)\,df.
\end{equation}
Afterwards we will take into account the fact that for modern cavity-locked lasers the cutoff frequency~$f_0$ between the white noise and~$1/f$ noise is $10\div100$~Hz~\cite{milani2017multiple, westergaard2010strontium}, which is much smaller than the width of the filter function~$\sim1/\tau$. Therefore, the integral can be divided into intervals, before and after~$f_0$:
\begin{equation}
\epsilon_T=\int\limits_{1/T_0}^{f_0}S_{\omega}(f)\mathcal{F}(f)\,df + \int\limits_{f_0}^{+\infty}S_{\omega}(f)\mathcal{F}(f)\,df\approx\mathcal{F}(0)\int\limits_{1/T_0}^{f_0}S_{\omega}(f)\,df + h_0\int\limits_{0}^{+\infty}\mathcal{F}(f)\,df.
\end{equation}
The second term is the error from the white noise approximation that was calculated before. The integral in the first term is the mean squared frequency deviation~\cite{riehle_frequency_2006}. Therefore, for the error induced by a low-frequency~$1/f$-type noise component, we get
\begin{equation}
\epsilon_{\mathrm{low}}\approx\mathcal{F}_{\mathrm{slow}}(0)\langle\delta\omega^2\rangle,
\end{equation}
where~$\delta\omega^2$ is the squared frequency deviation. Physically, this result can be understood as follows. The low-noise component is only prevalent at times much longer than the gate time. Therefore, during the gate the laser spectrum will behave like a narrow line that has the linewidth~$\Delta\omega$ and is shifted by a constant~$\delta\omega$. Thus, we can treat this constant shift as a noise that has a delta-like spectrum~$S_{\omega}(f) = \delta\omega^2\,\delta(f)$ during the gate. Then the convolution with the same filter function~$\mathcal{F}_{\mathrm{slow}}$ will yield the formula above. The averaging time~$T_0$ depends on the design of the experiment and should take the value that makes the most sense in each specific case. For example, this can be the total duration of an uninterrupted experimental sequence, where multiple runs of the same quantum circuit are executed without intermittent re-calibrations.}

\textcolor{black}{To summarize our findings, we estimate some experimental requirements that should be put onto the laser system for it to be able to perform useful quantum computation. The widely accepted error value for error correction to begin offering advantage is around~$10^{-3}$~\cite{ryan2021realization, bluvstein2024logical}. We take the same gate parameters that we used to plot the filter functions, namely: gate time~$\tau=167\ \mu\text{s}$, Rabi frequency~$\Omega = 2\pi\cdot 20\ \text{kHz}$. For this set of parameters, we obtain the following:
\begin{itemize}
    \item Phase noise PSD at~$\mu/2\pi$ $S_{\phi}(\mu/2\pi)<-93\ \text{dBc/Hz}$;
    \item Laser linewidth $\Delta\omega<2\ \text{Hz}$;
    \item Central frequency deviation $\delta\omega<65\ \text{Hz}$.
\end{itemize}
These characteristics are achievable in practice, but are quite demanding. They require long-term stability of the cavity (both thermal and mechanical), as well as a sufficiently low Schawlow-Townes white noise floor, which motivates the use of solid-state lasers over diodes. Furthermore, a -93~dBc/Hz level of noise at the detuning can only be realized using auxiliary techniques, such as implementing a filtering cavity~\cite{krinner2024low, semenin2025improved} or injection locking~\cite{krinner2024low, galstyan2024injection}.}




\vspace{6pt} 





\authorcontributions{
Conceptualization, N.S., K.K., N.K.; formal analysis, N.S.; investigation, N.S.; software, N.S.; visualization, N.S.; writing---original draft preparation, N.S.; writing---review and editing, K.K., N.K.; supervision, K.K., N.K. All authors have read and agreed to the published version of the manuscript.
}

\funding{This research was funded by the Russian State Corporation "Rosatom" within the Roadmap on Quantum Computing, Contract No. 868/1653-D dated 21 August 2025.}




\dataavailability{The original contributions presented in this study are included in the article. Further inquiries can be directed to the corresponding author(s).}

\conflictsofinterest{The authors declare no conflicts of interest.} 



\abbreviations{Abbreviations}{
The following abbreviations are used in this manuscript:
\\

\noindent 
\begin{tabular}{@{}ll}
MS gate & M\o lmer-S\o rensen gate\\
PSD & Power spectral density\\
QCCD & Quantum charge-coupled device\\
FWHM & Full width at half-maximum
\end{tabular}
}

\appendixtitles{yes} 
\appendixstart
\appendix
\section[\appendixname~\thesection]{Derivation of the formula for the mean gate error}
\label{app:mean_error}
This section is devoted to the derivation of equation~\eqref{eq:error_average} for the gate error averaged over all starting states, when the ideal and actual gate unitaries are~$\widehat{U}_0$ and~$\widehat{U}=\widehat{U}_0(1-i\widehat{T})$, respectively. We start from the formula for the average fidelity~\cite{nielsen_simple_2002, wu_noise_2018}:
\begin{multline}
\label{eq:fidelity_mean_full}
F=\dfrac{\sum\limits_n\Tr\left(\widehat{U}_0\widehat{W}^{\dagger}_n\widehat{U}^{\dagger}_0\Tr_m\left[\widehat{U}\widehat{W}_n\otimes\rho_m\widehat{U}^{\dagger}\right]\right) + d^2}{d^2(d+1)} \\
= \dfrac{\sum\limits_n\Tr\left(\widehat{W}^{\dagger}_n\Tr_m\left[(1-i\widehat{T})\widehat{W}_n\otimes\rho_m(1+i\widehat{T}^{\dagger})\right]\right) + d^2}{d^2(d+1)},
\end{multline}
where~$d=4$ for two qubits,~$\Tr_m$ is the trace over the motional degrees of freedom,~$\rho_m$ is the motional density matrix (assumed to be diagonal), and~$\widehat{W}_n$ are the orthogonal unitary operators constituting a basis, such that
\begin{equation}
\Tr(\widehat{W}_m^{\dagger}\widehat{W}_n) = d\delta_{mn}.
\end{equation}
We take a specific basis of operators as in~\cite{nielsen_simple_2002}:
\begin{equation}
\label{eq:W_n_definition}
\widehat{W}_n = X^kZ^l,
\end{equation}
where~$n=(k,l)$ is a multi-index with~$k,l=0,\dots,d-1$, and~$X$ and~$Z$ are the generalized Pauli matrices for a~$d$-level system that act as
\begin{equation}
\begin{aligned}
X\ket{n} &= \ket{n\oplus1},\\
Z\ket{n} &= e^{2\pi in/d}\ket{n},
\end{aligned}
\end{equation}
where~$\oplus$ is the addition modulo~$d$. Before simplifying~\eqref{eq:fidelity_mean_full}, we consider the sum
\begin{equation}
\sum\limits_n\widehat{W}_n^{\dagger}\widehat{W}_m\widehat{W}_n.
\end{equation}
Substituting~\eqref{eq:W_n_definition} with $n=(k, l), m = (k',l')$ and using the easily checked identity
\begin{equation}
Z^lX^k = e^{\frac{2\pi i}{d} kl}X^kZ^l,
\end{equation}
we obtain
\begin{equation}
\widehat{W}_n^{\dagger}\widehat{W}_m\widehat{W}_n = e^{\frac{2\pi i}{d}(kl'-lk')}\widehat{W}_m.
\end{equation}
Therefore, the sum will be
\begin{equation}
\label{eq:W_sum_trick}
\sum\limits_n\widehat{W}_n^{\dagger}\widehat{W}_m\widehat{W}_n = \sum\limits_k e^{\frac{2\pi i}{d}kl'}\sum\limits_l e^{-\frac{2\pi i}{d}lk'}\widehat{W}_m = d^2\delta_{m,0}\widehat{W}_m = d^2\delta_{m,0} I,
\end{equation}
where~$I$ is the identity operator.

The trace over motional variables in~\eqref{eq:fidelity_mean_full} is expanded as
\begin{equation}
\label{eq:trace_sum_aux}
\widehat{W}_n - i\Tr_m[\widehat{T}\widehat{W}_n\otimes\rho_m] + i\Tr_m[\widehat{W}_n\otimes\rho_m\widehat{T}^{\dagger}] + \Tr_m[\widehat{T}\widehat{W}_n\otimes\rho_m\widehat{T}^{\dagger}].
\end{equation}
The first term will contribute to the sum in~\eqref{eq:fidelity_mean_full} the following value:
\begin{equation}
\label{eq:first_trace_simplify}
\sum\limits_n \Tr(\widehat{W}_n^{\dagger}\widehat{W}_n) = d^3.
\end{equation}
The motional trace in the second term can be expressed as
\begin{equation}
\label{eq:TW_n_rho_aux}
\Tr_m[\widehat{T}\widehat{W}_n\otimes\rho_m] = \Tr_m[\widehat{T}\rho_m]\widehat{W}_n = \widehat{A}\widehat{W}_n,
\end{equation}
where~$\widehat{A}$ is the operator acting in the two-qubit space, therefore, it can be expanded in the basis~$\widehat{W}_k$:
\begin{equation}
\widehat{A} = \sum\limits_k\dfrac{\Tr(\widehat{W}_k^{\dagger}\widehat{A})}{\Tr(\widehat{W}_k^{\dagger}\widehat{W}_k)}\widehat{W}_k=\dfrac{1}{d}\sum\limits_k\Tr(\widehat{W}_k^{\dagger}\widehat{A})\widehat{W}_k.
\end{equation}
When multiplied by~$\widehat{W}_n^{\dagger}$, traced and summed over~$n$, the expression~\eqref{eq:TW_n_rho_aux} will be
\begin{multline}
\label{eq:trace_sum_simplify}
\sum\limits_n\Tr(\widehat{W}_n^{\dagger}\widehat{A}\widehat{W}_n) = \dfrac{1}{d}\sum\limits_n\sum\limits_k\Tr(\widehat{W}_k^{\dagger}\widehat{A})\Tr(\widehat{W}_n^{\dagger}\widehat{W}_k\widehat{W}_n) = d\sum\limits_k\Tr(\widehat{W}_k^{\dagger}\widehat{A})\delta_{k,0}\Tr I \\
= d^2\Tr\widehat{A} = d^2\Tr^{(0)}[\widehat{T}\rho_m],
\end{multline}
where~$\Tr^{(0)}$ is the full trace (over all variables), so the operators within this trace can be reordered: $\Tr^{(0)}[\widehat{T}\rho_m] = \Tr^{(0)}[\rho_m\widehat{T}]$. The third term in~\eqref{eq:trace_sum_aux} can be simplified likewise. Therefore, these two terms will contribute to the sum in~\eqref{eq:fidelity_mean_full} the following:
\begin{equation}
-id^2\Tr^{(0)}[\rho_m\widehat{T}] + id^2\Tr^{(0)}[\rho_m\widehat{T}^{\dagger}] = d^2\Tr^{(0)}[\rho_mi(\widehat{T}^{\dagger} - \widehat{T})].
\end{equation}
Since~$\widehat{U}=\widehat{U}_0(1-i\widehat{T})$ is unitary, it can be shown that
\begin{equation}
i(\widehat{T}^{\dagger} - \widehat{T}) = -\widehat{T}^{\dagger}\widehat{T}.
\end{equation}
Therefore, the contribution to the sum in~\eqref{eq:fidelity_mean_full} will be
\begin{equation}
d^2\Tr^{(0)}[\rho_mi(\widehat{T}^{\dagger} - \widehat{T})] = -d^2\Tr^{(0)}[\rho_m(\widehat{T}^{\dagger}\widehat{T})] = -d^2\sum\limits_n p_n \left[\Tr(\widehat{T}^{\dagger}\widehat{T})\right]_{nn},
\end{equation}
since~$\rho_m$ is diagonal. The last term in~\eqref{eq:trace_sum_aux} is
\begin{equation}
\Tr_m[\widehat{T}\widehat{W}_n\otimes\rho_m\widehat{T}^{\dagger}] = \sum\limits_{mk}p_k\widehat{T}_{mk}\widehat{W}_n\widehat{T}^{\dagger}_{km}.
\end{equation}
Multiplying by~$\widehat{W}_n^{\dagger}$, tracing and summing over~$n$, we obtain, analogously to~\eqref{eq:trace_sum_simplify}:
\begin{multline}
\label{eq:last_trace_simplify}
\sum\limits_{mk}p_k\Tr\left(\sum\limits_{n}\widehat{W}_n^{\dagger}\widehat{T}_{mk}\widehat{W}_n\widehat{T}^{\dagger}_{km}\right) = d\sum\limits_{mk}p_k\Tr\left(\left(\Tr\widehat{T}_{mk}\right)\widehat{T}^{\dagger}_{km}\right) = d\sum\limits_{mk}p_k\left(\Tr\widehat{T}_{mk}\right)\Tr\left(\widehat{T}^{\dagger}_{km}\right) \\
= d\sum\limits_{mk}p_k\left|\Tr\widehat{T}_{mk}\right|^2.
\end{multline}
Combining~\eqref{eq:first_trace_simplify},~\eqref{eq:trace_sum_simplify},~\eqref{eq:last_trace_simplify} and substituting into~\eqref{eq:fidelity_mean_full}, we obtain the average gate error:
\begin{equation}
\epsilon=1-F= \dfrac{1}{d+1} \sum\limits_n p_n \left[\Tr(\widehat{T}^{\dagger}\widehat{T})\right]_{nn} - \dfrac{1}{d(d+1)}\sum\limits_{mn} p_m \left|(\Tr\widehat{T})_{nm}\right|^2.
\end{equation}
For two qubits~$d=4$, and
\begin{equation}
\epsilon = \dfrac{1}{5}\sum\limits_n p_n \left[\Tr(\widehat{T}^{\dagger}\widehat{T})\right]_{nn} - \dfrac{1}{20}\sum\limits_{mn} p_m \left|(\Tr\widehat{T})_{nm}\right|^2,
\end{equation}

\begin{adjustwidth}{-\extralength}{0cm}

\reftitle{References}


\bibliography{bibliography.bib}

\PublishersNote{}
\end{adjustwidth}
\end{document}